\documentclass[
reprint,
aps,
prd, 
superscriptaddress,
showpacs,
nofootinbib,
amsmath,
amssymb,
longbibliography 
]{revtex4-2}
\usepackage{slashed}
\usepackage{graphicx}
\usepackage{dcolumn}
\usepackage{bm}
\usepackage{hyperref}
\usepackage{color}
\usepackage{orcidlink}
\usepackage{filecontents}
\usepackage{float}
\usepackage{booktabs}
\usepackage{mathtools}
\usepackage{tikz}

\usetikzlibrary{arrows.meta, positioning, shapes.geometric, calc, fit, backgrounds}

\usepackage{xcolor}
\definecolor{darkblue}{rgb}{0.0, 0.0, 0.55}
\definecolor{darkred}{rgb}{0.55, 0.0, 0.0}

\hypersetup{
	colorlinks=true,
	linkcolor=darkblue, 
	citecolor=darkred,  
	urlcolor=darkblue   
}

\begin{document}
	
	\title{External-Field-Assisted Muon Reactivation in Muon-Catalyzed Fusion: A Rate-Network Criterion for Reducing Alpha Sticking}
	
	\author{Wei Kou\orcidlink{0000-0002-4152-2150}}
	\email{kouwei@impcas.ac.cn}
	\affiliation{Institute of Modern Physics, Chinese Academy of Sciences, Lanzhou 730000, Gansu Province, China}
	\affiliation{Southern Center for Nuclear Science Theory (SCNT), Institute of Modern Physics, Chinese Academy of Sciences, Huizhou 516000, Guangdong Province, China}
	\affiliation{School of Nuclear Science and Technology, University of Chinese Academy of Sciences, Beijing 100049, China}
	\affiliation{State Key Laboratory of Heavy Ion Science and Technology, Institute of Modern Physics, Chinese Academy of Sciences, Lanzhou 730000, Gansu Province, China}
	
	\author{Xurong Chen}
	\email{xchen@impcas.ac.cn}
	\affiliation{Institute of Modern Physics, Chinese Academy of Sciences, Lanzhou 730000, Gansu Province, China}
	\affiliation{Southern Center for Nuclear Science Theory (SCNT), Institute of Modern Physics, Chinese Academy of Sciences, Huizhou 516000, Guangdong Province, China}
	\affiliation{School of Nuclear Science and Technology, University of Chinese Academy of Sciences, Beijing 100049, China}
	\affiliation{State Key Laboratory of Heavy Ion Science and Technology, Institute of Modern Physics, Chinese Academy of Sciences, Lanzhou 730000, Gansu Province, China}

	\begin{abstract}
		Alpha sticking is a major loss channel in deuterium--tritium
		muon-catalyzed fusion.  We study whether an additional
		external-field-assisted stripping channel can reduce the residual sticking
		loss after conventional collisional reactivation.  The external contribution
		is written as
		\(R_X=f_XP_X\eta_X\), where \(f_X\) is the space--time overlap between the
		external field and the residual stuck \((\alpha\mu)^+\) population, \(P_X\)
		is the microscopic stripping probability, and \(\eta_X\) is the probability
		that the stripped \(\mu^-\) is returned to the \(d\mu/t\mu\to dt\mu\)
		fusion cycle before escape or decay.  This gives
		\(\omega_S^{\rm eff}=\omega_S^0(1-R_{\rm col})(1-R_X)\) and leads directly
		to a probability-level no-go condition, \(\eta_X^{\rm crit}>1\), for any
		target improvement requiring more recycling than is probabilistically
		available. We construct an energy-resolved post-stripping rate network including
		slowing down, atomic capture, free escape, muon decay, atomic-stage loss,
		ordinary molecular formation, and an effective resonant \(dt\mu\) channel.
		Benchmark scans show that the useful regime is a transport window: the
		stripped muon must be confined and recycled efficiently.  With the reference
		inputs used here, the best-performing scenario increases the cycle yield from
		\(N_{\rm fus,\mu}=112.6\) in the collision-only case to
		\(N_{\rm fus,\mu}=156.5\).  Resonant molecular formation suppresses
		atomic-stage loss and broadens the high-recycling region, but it cannot
		compensate for prompt escape or poor field--population overlap.  The rate
		network therefore identifies the transport and overlap conditions required
		for external-field-assisted reactivation to reduce residual alpha sticking.
	\end{abstract}

	
	\maketitle

	\section{Introduction}
	\label{sec:introduction}
	
	Muon-catalyzed fusion (\(\mu\)CF) is a low-temperature fusion mechanism in
	which a negative muon replaces an electron and compresses the molecular length
	scale by approximately the muon-to-electron mass ratio.  In a
	hydrogen-isotope target, the muon first forms a muonic atom and can then form
	a muonic molecule, bringing the nuclei close enough for fusion through
	Coulomb-barrier tunneling.  The idea dates back to the early development of
	muon physics
	\cite{frank1947hypothetical,jackson1957catalysis,Alvarez:1957un}
	and has since been studied as a few-body reaction problem, a source of fusion
	neutrons, and a possible route toward fusion-energy applications
	\cite{ponomarev1990muon,froelich1992muon,Petitjean:1992iq}.
	
	Among hydrogen-isotope systems, the deuterium--tritium channel gives the
	largest \(\mu\)CF yield.  The ground-state \(dt\mu\) molecule undergoes
	\begin{equation}
		(dt\mu)_{J=v=0}
		\rightarrow
		\alpha+n+\mu^-+17.6~{\rm MeV},
		\label{eq:intro_dtmu_free}
	\end{equation}
	after which the released muon can, in principle, catalyze another fusion
	event.  A competing channel is
	\begin{equation}
		(dt\mu)_{J=v=0}
		\rightarrow
		(\alpha\mu)_{n\ell}+n+17.6~{\rm MeV},
		\label{eq:intro_dtmu_sticking}
	\end{equation}
	where the muon remains bound to the outgoing alpha particle.  This
	alpha-sticking process removes the muon from the catalytic cycle unless the
	bound muon is subsequently stripped from the \((\alpha\mu)^+\) ion in the
	target medium.
	
	The number of fusion reactions catalyzed by one muon is mainly limited by
	muon decay and residual alpha sticking.  The usual kinetic description
	separates the initial sticking probability \(\omega_S^0\) from the
	reactivation probability \(R\), giving
	\begin{equation}
		\omega_S^{\rm eff}
		=
		\omega_S^0(1-R).
		\label{eq:intro_standard_sticking}
	\end{equation}
	The corresponding cycle estimate is
	\begin{equation}
		N_{\rm fus,\mu}
		\simeq
		\frac{\phi\lambda_c}
		{
			\lambda_\mu+\omega_S^{\rm eff}\phi\lambda_c
		},
		\label{eq:intro_cycle_number}
	\end{equation}
	where \(\phi\) is the target density relative to liquid hydrogen,
	\(\lambda_c\) is the effective cycle rate, and
	\(\lambda_\mu=\tau_\mu^{-1}\) is the muon decay rate.  This expression shows
	why even a sub-percent residual sticking probability can strongly reduce the
	cycle yield after many catalytic turns.
	
	Recent few-body calculations have sharpened the microscopic input to this
	problem.  Kamimura, Kino, and Yamashita solved the coupled-channel
	three-body reaction
	\[
	(dt\mu)_{J=v=0}
	\rightarrow
	\alpha+n+\mu^-+17.6~{\rm MeV}
	\]
	and
	\[
	(dt\mu)_{J=v=0}
	\rightarrow
	(\alpha\mu)_{n\ell}+n+17.6~{\rm MeV},
	\]
	obtaining the fusion rate, the transition rates to bound and continuum
	\(\alpha\mu\) states, the initial sticking probability, and the emitted-muon
	spectrum in a unified framework \cite{Kamimura:2021msf}.  Wu and Kamimura
	then developed a tractable \(T\)-matrix formulation that reproduces the main
	coupled-channel results and can be extended to other muonic molecules and
	reaction channels \cite{Wu:2024uad}.  These studies provide the microscopic
	sticking and fusion inputs used below.
	
	At the kinetic level, resonant muonic molecules can modify the \(\mu\)CF
	cycle by changing isotope populations, producing epithermal muonic atoms, and
	opening fast molecular-formation pathways
	\cite{iiyoshi2019muon,Yamashita:2022rtu}.  Radiative-decay calculations for
	\(dd\mu^\ast\) and \(dt\mu^\ast\) resonances further suggest direct routes
	to bound-state muonic molecules \cite{yamashita2025radiative}.  The direct
	observation of muonic molecules in resonance states by high-resolution x-ray
	spectroscopy has also strengthened the experimental basis for studying such
	channels \cite{toyama2026direct}.  These developments make the reactivation
	stage, not only the molecular-formation stage, worth revisiting.
	
	External fields provide a possible handle on alpha sticking.  Kimura and
	Bonasera studied an x-ray-laser-assisted \(d\)-\(t\) \(\mu\)CF scenario with
	emphasis on the alpha-sticking problem \cite{kimura2008application}.  Related
	laser-assisted mechanisms have been considered for in-flight muon-catalyzed
	deuteron--triton fusion, where an intense laser field modifies the collision
	dynamics of a deuteron with a muonic tritium atom \cite{Liu:2022gbs}.  Mori
	proposed a more targeted reactivation mechanism based on resonance
	radio-frequency acceleration of \(\mu{\rm He}^+\) ions, aiming to enhance
	muon stripping from the alpha particle and reduce the effective sticking loss
	\cite{mori2021enforced}.  These studies show that external fields can act on
	specific stages of the \(\mu\)CF cycle.  The remaining issue is kinetic
	closure: a stripped muon contributes to the cycle only if it slows down,
	is captured, forms \(dt\mu\), and undergoes fusion before escape or decay.
	
	In this work we formulate this closure as a rate-network criterion for
	external-field-assisted muon reactivation.  Conventional collisional
	reactivation is taken as the baseline, and the externally induced branch is
	written in terms of three factors: the field--population overlap, the
	microscopic stripping probability, and the post-stripping recycling
	probability.  The post-stripping dynamics includes slowing down, atomic
	capture, free escape, muon decay, atomic-stage loss, ordinary molecular
	formation, and an effective resonant \(dt\mu\) fast channel.  The resulting
	framework identifies the transport and overlap conditions required for an
	external channel to reduce the residual alpha-sticking loss.
	
	The paper is organized as follows.  In Sec.~\ref{sec:framework}, we introduce
	the physical picture and derive the combined reactivation criterion.  In
	Sec.~\ref{sec:rate_network}, we construct the post-stripping rate network for
	the liberated muon.  In Sec.~\ref{sec:numerical_setup}, we specify the
	numerical implementation, benchmark inputs, and parameter scans.  In
	Sec.~\ref{sec:results}, we present the benchmark scenarios, transport-window
	maps, the no-go condition, and the robustness tests.  Finally, the main
	conclusions are summarized in Sec.~\ref{sec:conclusion}.

	\section{Physical picture and reactivation criterion}
	\label{sec:framework}
	
	We separate the short-time sticking process from the subsequent reactivation
	dynamics.  The initial sticking probability \(\omega_S^0\) is fixed by the
	branching of the \(dt\mu\) fusion reaction into the continuum
	\(\alpha+n+\mu^-\) channel and the bound \((\alpha\mu)_{n\ell}+n\) channel.
	Recent few-body calculations determine this quantity with improved treatment
	of both bound and continuum \(\alpha\mu\) final states
	\cite{Kamimura:2021msf,Wu:2024uad}.  In the following, \(\omega_S^0\) is
	taken as an input, and we focus on the reactivation of the residual
	\((\alpha\mu)^+\) population.
	
	For the collision-only baseline, the reactivation probability is denoted by
	\(R_{\rm col}\).  The corresponding residual sticking probability is
	\begin{equation}
		\omega_{S,{\rm col}}^{\rm eff}
		=
		\omega_S^0(1-R_{\rm col}) .
		\label{eq:omega_col_only}
	\end{equation}
	Here \(R_{\rm col}\) summarizes conventional collisional stripping in the
	D--T medium.
	
	We now add an externally driven stripping branch.  For definiteness, we use
	the language of x-ray photostripping of the bound \(\alpha\mu\) state, but
	the same criterion applies to any external mechanism that liberates the muon
	from \((\alpha\mu)^+\).  The net external reactivation probability is written
	as
	\begin{equation}
		R_X
		=
		f_X P_X \eta_X .
		\label{eq:RX_definition}
	\end{equation}
	Here \(P_X\) is the microscopic stripping probability, \(f_X\) is the
	space--time overlap between the external field and the residual stuck
	population, and \(\eta_X\) is the probability that the stripped \(\mu^-\) is
	returned to the \(d\mu/t\mu\to dt\mu\) fusion cycle before escape or decay.
	
\begin{figure}[htbp]
		\centering
		\includegraphics[width=0.49\textwidth]{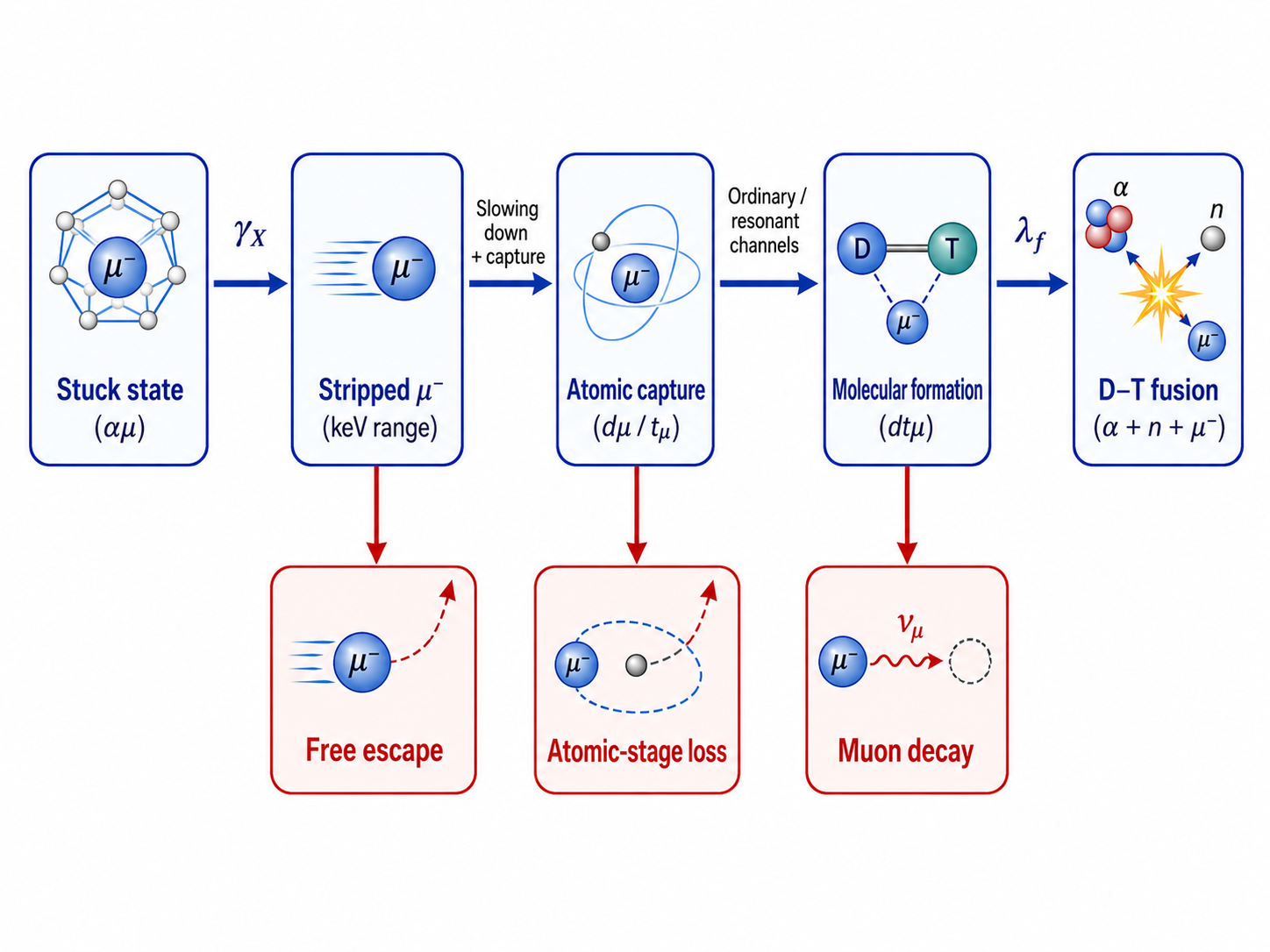}
		\caption{
			External-field-assisted reactivation pathway.  The external field strips
			a residual \((\alpha\mu)^+\) state and produces a free \(\mu^-\).  The
			muon contributes to the catalytic cycle only if it remains confined,
			slows down, is captured into the \(d\mu/t\mu\) atomic stage, forms
			\(dt\mu\), and undergoes fusion.  Escape, atomic-stage loss, and muon
			decay compete with this recycling branch.
		}
		\label{fig:reactivation_schematic}
	\end{figure}

	The overlap factor may be defined as
	\begin{equation}
		\begin{aligned}
			f_X
			&=
			\int d^3{\bf r}\,dt\,
			p_{\rm res}({\bf r},t)\,
			W_X({\bf r},t),
			\\
			1
			&=
			\int d^3{\bf r}\,dt\,
			p_{\rm res}({\bf r},t),
			\qquad
			0\leq W_X({\bf r},t)\leq 1 .
		\end{aligned}
		\label{eq:fX_overlap_definition}
	\end{equation}
	Here \(p_{\rm res}({\bf r},t)\) is the normalized distribution of residual
	\((\alpha\mu)^+\) ions after collisional reactivation, and \(W_X\) describes
	the external pulse window.  Thus \(f_X\) includes both geometrical coverage
	and timing relative to the surviving stuck-ion population. The corresponding recycling pathway is sketched in Fig.~\ref{fig:reactivation_schematic}.
	
	Combining conventional collisional reactivation with the external branch gives
	\begin{equation}
		\omega_S^{\rm eff}
		=
		\omega_S^0(1-R_{\rm col})(1-R_X).
		\label{eq:omega_combined}
	\end{equation}
	Equation~\eqref{eq:omega_combined} is exact in the sequential limit where
	collisional reactivation first removes a fraction \(R_{\rm col}\), and the
	external field then acts independently on the remaining population.  If the
	two mechanisms overlap in time or act on correlated subensembles,
	Eq.~\eqref{eq:omega_combined} should be read as a factorized survival
	approximation, with \(R_X\) representing the additional reactivation not
	already included in \(R_{\rm col}\).  This form is sufficient for the
	probability-level criterion developed here; a fully time-dependent treatment
	would require coupled rate equations for the residual \((\alpha\mu)^+\)
	population, collisional stripping, external stripping, post-stripping
	transport, and recycling.
	
	The cycle-yield gain is obtained by substituting
	Eq.~\eqref{eq:omega_combined} into Eq.~\eqref{eq:intro_cycle_number}.  We use
	\begin{equation}
		G_N
		=
		\frac{
			N_{\rm fus,\mu}(R_X)
		}{
			N_{\rm fus,\mu}(R_X=0)
		},
		\label{eq:GN_definition}
	\end{equation}
	where the denominator is the collision-only baseline at fixed
	\(\omega_S^0\), \(R_{\rm col}\), \(\phi\), and \(\lambda_c\).
	
	The same factorization gives a simple no-go condition.  If a target
	improvement requires \(R_X>R_X^{\rm crit}\), then the required recycling
	probability is
	\begin{equation}
		\eta_X^{\rm crit}
		=
		\frac{R_X^{\rm crit}}{f_XP_X}.
		\label{eq:eta_crit}
	\end{equation}
	Since \(\eta_X\leq 1\), any parameter region satisfying
	\begin{equation}
		\eta_X^{\rm crit}>1
		\label{eq:no_go_condition}
	\end{equation}
	is excluded independently of the detailed transport model.  The rate-network
	calculation below therefore determines whether the post-stripping dynamics
	can provide an \(\eta_X\) large enough to satisfy
	Eq.~\eqref{eq:eta_crit}.

	\section{Rate-network model}
	\label{sec:rate_network}
	
	The criterion in Sec.~\ref{sec:framework} reduces the external channel to
	three factors: the field--population overlap \(f_X\), the microscopic
	stripping probability \(P_X\), and the post-stripping recycling probability
	\(\eta_X\).  The first two factors determine how many residual
	\((\alpha\mu)^+\) ions produce stripped muons.  The last factor describes
	whether a liberated \(\mu^-\) slows down, is captured into the \(d\mu/t\mu\)
	atomic stage, forms \(dt\mu\), and reaches fusion before escape, decay, or
	atomic-stage loss.  This section defines the energy-resolved rate network
	used to compute \(\eta_X\).
	
	\subsection{Post-stripping spectrum and effective inputs}
	\label{subsec:spectrum_inputs}
	
	For a given external photon field, the stripped muon is described by an
	energy distribution \(F_X(E)\), normalized as
	\begin{equation}
		\int_0^\infty F_X(E)\,dE = 1 .
		\label{eq:FX_norm}
	\end{equation}
	As a reference kinematic estimate, consider an isolated ground-state
	\(\alpha\mu\) atom initially at rest.  The central kinetic energy of the
	emitted muon is then
	\begin{equation}
		E_\mu^{(0)}
		=
		\frac{m_\alpha}{m_\alpha+m_\mu}
		\left(\hbar\omega_X-\left|E_{1s}^{\alpha\mu}\right|\right),
		\qquad
		\hbar\omega_X>\left|E_{1s}^{\alpha\mu}\right| ,
		\label{eq:stripped_energy_center}
	\end{equation}
	where \(E_{1s}^{\alpha\mu}\) is the ground-state binding energy of
	\(\alpha\mu\).  This is a rest-frame expression.  In a target medium, the
	residual \((\alpha\mu)^+\) ion may retain a center-of-mass velocity
	\({\bf V}_{\alpha\mu}\) from the fusion recoil and from subsequent slowing
	collisions.  If the muon is emitted with rest-frame kinetic energy
	\(E_\mu^\ast\) and direction \(\hat{\bf n}\), its nonrelativistic
	laboratory-frame energy is
	\begin{equation}
		E_{\mu}^{\rm lab}
		=
		E_\mu^\ast
		+
		\frac{1}{2}m_\mu V_{\alpha\mu}^2
		+
		\sqrt{2m_\mu E_\mu^\ast}\,
		V_{\alpha\mu}\cos\theta ,
		\label{eq:boosted_muon_energy}
	\end{equation}
	with \(\cos\theta=\hat{\bf n}\cdot\hat{\bf V}_{\alpha\mu}\).  The effective
	stripped-muon spectrum is therefore a convolution over the residual ion
	velocity distribution,
	\begin{equation}
		F_X(E)
		=
		\int d^3{\bf V}_{\alpha\mu}\,
		G_{\alpha\mu}({\bf V}_{\alpha\mu};t_X)
		\int_{-1}^{1}\frac{d\cos\theta}{2}\,
		\delta\!\left(E-E_{\mu}^{\rm lab}\right),
		\label{eq:FX_boost_convolution}
	\end{equation}
	with additional broadening from pulse bandwidth and target-medium effects.
	The slow-stuck limit,
	\(G_{\alpha\mu}({\bf V})\rightarrow \delta^3({\bf V})\), reduces
	Eq.~\eqref{eq:FX_boost_convolution} to
	Eq.~\eqref{eq:stripped_energy_center}.  The benchmark calculations use this
	limit as the reference spectrum.  Irradiation of a fast residual
	\((\alpha\mu)^+\) component would require the boosted spectrum in
	Eq.~\eqref{eq:FX_boost_convolution}.  The width and shape of \(F_X(E)\) are
	varied in Sec.~\ref{sec:results} to test the sensitivity to the assumed
	spectrum.
	
	The remaining transport inputs are the effective capture cross section, the
	stopping power, the confinement length, and the resonant molecular-formation
	rate.  These quantities are treated as scan parameters in the present
	criterion analysis; the network determines which combinations of them give a
	large enough post-stripping recycling probability.
	
	\subsection{Free-muon transport and atomic capture}
	\label{subsec:free_transport}
	
	After stripping, the free muon is propagated on an energy grid.  At kinetic
	energy \(E\), we use the nonrelativistic velocity
	\begin{equation}
		v_\mu(E)
		=
		c\sqrt{\frac{2E}{m_\mu c^2}},
		\label{eq:mu_velocity}
	\end{equation}
	which is adequate for the keV-scale energies considered here.  The effective
	capture rate into the muonic-atom stage is written as
	\begin{equation}
		\lambda_{\rm cap}(E)
		=
		n_{D/T}\,
		\sigma_{\rm cap}^{\rm eff}(E)\,
		v_\mu(E),
		\label{eq:lambda_cap}
	\end{equation}
	where \(n_{D/T}=\phi n_{\rm LH}\).  This is the standard kinetic-theory
	collision-rate estimate \(n\sigma v\) \cite{chapman1970mathematical}.  Here
	\(\sigma_{\rm cap}^{\rm eff}(E)\) is an effective capture cross section that
	coarse-grains atomic capture, target-medium effects, cascade processes, and
	other short-distance details not resolved in the network.
	
	Continuous slowing down is represented by bin-to-bin transitions on the
	energy grid.  We define
	\begin{equation}
		S_{\rm eff}(E)
		=
		-\frac{dE}{dx},
		\label{eq:stopping_power_definition}
	\end{equation}
	following the standard charged-particle stopping-power convention
	\cite{bethe1930theorie}.  The corresponding energy-loss rate is
	\(\left|dE/dt\right|=S_{\rm eff}(E)v_\mu(E)\).  For adjacent energy bins,
	the discrete slowing rate is
	\begin{equation}
		\lambda_{\rm slow}(E_i)
		=
		\frac{\left|dE/dt\right|_{E_i}}{\Delta E_i},
		\qquad
		\left|\frac{dE}{dt}\right|_{E_i}
		=
		S_{\rm eff}(E_i)v_\mu(E_i),
		\label{eq:lambda_slow}
	\end{equation}
	where \(\Delta E_i=E_i-E_{i-1}\).  Thus
	\(\lambda_{\rm slow}\) is not an independent microscopic reaction rate, but
	the discrete representation of continuous energy loss.  The dependence on
	the energy grid is checked in Sec.~\ref{sec:results}.
	
	Loss from the active target region is described by an escape-time
	approximation.  If a muon with velocity \(v_\mu(E)\) samples an effective
	confinement length \(L_{\rm eff}\), its residence time is of order
	\(L_{\rm eff}/v_\mu(E)\).  We write
	\begin{equation}
		\lambda_{\rm esc}(E)
		=
		\kappa_{\rm esc}\frac{v_\mu(E)}{L_{\rm eff}},
		\label{eq:lambda_escape}
	\end{equation}
	where \(\kappa_{\rm esc}\) accounts for geometry, angular distribution,
	scattering, and partial confinement.  This finite-volume loss closure is
	analogous to inverse-residence-time descriptions used in transport and
	reactor-flow modeling \cite{danckwerts1953continuous}.
	
	\subsection{Molecular formation, recycling probability, and loss budget}
	\label{subsec:recycling_budget}
	
	After capture, the muon enters an effective \(d\mu/t\mu\) atomic stage.  From
	there it may form \(dt\mu\), be lost in atomic-stage channels, decay, or
	escape from the active region.  The net molecular-formation rate is
	parametrized as
	\begin{equation}
		\lambda_{dt\mu}^{\rm eff}
		=
		\lambda_{dt\mu}^{\rm ord}
		+
		\lambda_{dt\mu}^{\rm res}.
		\label{eq:lambda_dtmu_eff}
	\end{equation}
	Here \(\lambda_{dt\mu}^{\rm ord}\) denotes the ordinary molecular-formation
	contribution, while \(\lambda_{dt\mu}^{\rm res}\) represents an effective
	resonant fast channel.  Both are density-, mixture-, temperature-, and
	hyperfine-averaged rates.  This parametrization is motivated by recent
	studies of resonant muonic molecules in \(\mu\)CF kinetics
	\cite{Yamashita:2022rtu,yamashita2025radiative,toyama2026direct}.
	
	Once a fusion-active \(dt\mu\) molecule is formed, the intramolecular fusion
	rate is much larger than the free-muon decay rate.  We therefore treat
	\(\lambda_f\) as the dominant rate in the final fusion step and take it from
	few-body calculations \cite{Kamimura:2021msf,Wu:2024uad}.  The relevant
	losses in this network occur before the final fusion step: free escape after
	stripping, loss in the atomic stage, decay during transport, and failure to
	enter the fusion-active \(dt\mu\) channel.
	
	For each initial stripped-muon energy \(E\), the network returns the
	absorbing probabilities
	\begin{equation}
		P_{\rm fus}(E),\quad
		P_{\rm dec}(E),\quad
		P_{\rm esc}(E),\quad
		P_{\rm atom}(E),\quad
		P_{\rm form}(E),
		\label{eq:absorbing_probabilities}
	\end{equation}
	corresponding to successful fusion recycling, muon decay, free escape,
	atomic-stage loss, and failure to reach the fusion-active molecular channel.
	For each energy bin they satisfy
	\begin{equation}
		P_{\rm fus}(E)
		+
		P_{\rm dec}(E)
		+
		P_{\rm esc}(E)
		+
		P_{\rm atom}(E)
		+
		P_{\rm form}(E)
		=
		1 .
		\label{eq:energy_budget_sum}
	\end{equation}
	The absorbing-probability recursion used numerically follows directly from
	the competing rates defined above and propagates the probabilities across the
	energy grid.
	
	The post-stripping recycling probability in Eq.~\eqref{eq:RX_definition} is
	the spectrum-averaged successful-fusion probability,
	\begin{equation}
		\eta_X
		=
		\int_0^\infty
		F_X(E)\,P_{\rm fus}(E)\,dE .
		\label{eq:etaX_integral}
	\end{equation}
	The other loss components are averaged in the same way:
	\begin{align}
		P_{\rm dec}
		&=
		\int_0^\infty F_X(E)\,P_{\rm dec}(E)\,dE,
		\label{eq:Pdec_integral}
		\\
		P_{\rm esc}
		&=
		\int_0^\infty F_X(E)\,P_{\rm esc}(E)\,dE,
		\label{eq:Pesc_integral}
		\\
		P_{\rm atom}
		&=
		\int_0^\infty F_X(E)\,P_{\rm atom}(E)\,dE,
		\label{eq:Patom_integral}
		\\
		P_{\rm form}
		&=
		\int_0^\infty F_X(E)\,P_{\rm form}(E)\,dE.
		\label{eq:Pform_integral}
	\end{align}
	The spectrum-averaged budget is therefore
	\begin{equation}
		\eta_X
		+
		P_{\rm dec}
		+
		P_{\rm esc}
		+
		P_{\rm atom}
		+
		P_{\rm form}
		=
		1 .
		\label{eq:budget_sum}
	\end{equation}
	This normalization is used as an internal check and gives a channel-by-channel
	diagnosis of the external reactivation pathway.
	
	The network is intentionally minimal.  It does not replace microscopic
	calculations of the photostripping cross section
	\(\sigma_\gamma(\omega_X)\), the effective capture cross section
	\(\sigma_{\rm cap}^{\rm eff}(E)\), or the stopping power \(S_{\rm eff}(E)\).
	Those quantities are kept as scan inputs.  The purpose of the model is to
	connect microscopic stripping and capture inputs to the macroscopic question
	posed in Sec.~\ref{sec:framework}: whether post-stripping recycling can be
	large enough to reduce the residual sticking loss.

	\section{Numerical setup}
	\label{sec:numerical_setup}
	
	We now specify the numerical implementation of the criterion and rate network
	defined in Secs.~\ref{sec:framework} and \ref{sec:rate_network}.  The
	microscopic sticking and fusion inputs are fixed, while the capture, slowing,
	confinement, and resonant-formation parameters are varied to map the
	post-stripping transport response.
	
	The fixed microscopic inputs are taken from recent few-body calculations of
	the \(dt\mu\) fusion reaction.  We use
	\begin{equation}
		\omega_S^0 = 8.57\times 10^{-3},
		\qquad
		\lambda_f = 1.15\times 10^{12}~{\rm s}^{-1},
		\label{eq:fixed_micro_inputs}
	\end{equation}
	for the initial sticking probability and intramolecular fusion rate,
	respectively \cite{Kamimura:2021msf,Wu:2024uad}.  Thus the calculation below
	changes only the post-sticking reactivation and recycling part of the cycle.
	
	The free-muon decay rate is fixed by the PDG mean lifetime,
	\(\tau_\mu\simeq 2.2~\mu{\rm s}\), giving
	\begin{equation}
		\lambda_\mu
		=
		\tau_\mu^{-1}
		=
		4.55\times10^5~{\rm s}^{-1}.
	\end{equation}
	The target density is measured relative to the conventional liquid-hydrogen
	density used in \(\mu\)CF kinetics,
	\begin{equation}
		n_{\rm LH}
		=
		4.25\times10^{22}~{\rm cm}^{-3}.
	\end{equation}
	Unless otherwise stated, we use a dense D--T benchmark with
	\(\phi=1.25\) and an equal isotopic mixture, \(c_D=c_T=0.5\).  This density is
	within the range of high-density hydrogen-isotope targets explored in
	\(\mu\)CF studies, and the equal mixture isolates the post-stripping transport
	effect from isotope-composition effects
	\cite{ParticleDataGroup:2024cfk,Brunelli:1987MuonCatalyzed,Ackerbauer:1999,Bom:2005PAN}.
	
	For the collision-only reference case we take
	\begin{equation}
		R_{\rm col}
		=
		0.35,
	\end{equation}
	so that \(1-R_{\rm col}=0.65\).  This is used as a representative
	high-density collisional reactivation baseline, consistent with previous
	\(\mu\)CF reactivation estimates and recent few-body \(\mu\)CF calculations
	\cite{Rafelski:1989Reactivation,Kamimura:2021msf}.
	
	The external-field-induced stripping probability is evaluated with a
	photon-fluence model.  We use the threshold profile
	\begin{equation}
		\sigma_\gamma(\omega_X)
		=
		\sigma_{\rm pk}
		\exp\left[
		-\frac{(\hbar\omega_X-E_{\rm pk})^2}{2\Delta_X^2}
		\right]
		\Theta\!\left(\hbar\omega_X-\left|E_{1s}^{\alpha\mu}\right|\right),
		\label{eq:sigma_gamma_model}
	\end{equation}
	where the step function imposes the photostripping threshold.  The Gaussian
	factor is a phenomenological bandwidth profile, and \(\sigma_{\rm pk}\) is an
	effective strength used in the scan.  A microscopic calculation would require
	the bound--continuum transition matrix element of the \(\alpha\mu\) system
	and the corresponding continuum-state normalization
	\cite{fano1968spectral,bethe1957quantum}.
	
	For a photon fluence \(\Phi_X\), the mean number of independent stripping
	attempts is \(\sigma_\gamma(\omega_X)\Phi_X g_X\), where \(g_X\) is a
	geometrical factor.  The stripping probability is modeled as
	\begin{equation}
		P_X
		=
		1-\exp\!\left[-\sigma_\gamma(\omega_X)\Phi_X g_X\right].
		\label{eq:PX_fluence_model}
	\end{equation}
	This is the standard Poisson survival form, equivalent in structure to the
	Bouguer--Beer--Lambert attenuation law for independent absorption events
	\cite{mayerhofer2020bouguer}.
	
	The reference photon-field parameters are
	\begin{equation}
		\begin{aligned}
			\hbar\omega_X &= 15~{\rm keV},
			&
			\sigma_{\rm pk} &= 20~{\rm barn},
			\\
			E_{\rm pk} &= 15.5~{\rm keV},
			&
			\Delta_X &= 4~{\rm keV},
		\end{aligned}
	\end{equation}
	with
	\begin{equation}
		\Phi_X
		=
		3.0\times10^{22}~{\rm photons\,cm}^{-2},
		\qquad
		g_X
		=
		1 .
	\end{equation}
	These values give \(P_X\simeq0.449\).  The corresponding energy fluence is
	\begin{equation}
		\begin{aligned}
			{\cal E}_X
			&=
			\Phi_X\hbar\omega_X
			\\
			&\simeq
			7.2\times10^7~{\rm J\,cm}^{-2}
			\left(
			\frac{\Phi_X}{3.0\times10^{22}~{\rm cm}^{-2}}
			\right)
			\left(
			\frac{\hbar\omega_X}{15~{\rm keV}}
			\right).
			\label{eq:xray_energy_fluence}
		\end{aligned}
	\end{equation}
	This large fluence should be read as a benchmark input for the criterion
	calculation.  A concrete pulse design would also have to account for
	attainable fluence, focal volume, repetition rate, target damage, and
	synchronization with the residual \((\alpha\mu)^+\) population.  In the
	benchmark scenarios we set \(f_X=1\) to isolate the transport limitation; the
	dependence on \(f_X\) is studied separately in Sec.~\ref{subsec:overlap_nogo}
	\footnote{
		Here ``x-ray free-electron laser'' (XFEL) is used only as a representative
		example of an intense pulsed x-ray external field.  The criterion is general
		and applies to any external stripping mechanism parametrized by \(f_X\),
		\(P_X\), and \(\eta_X\).
	}.
	
	The stripped-muon spectrum is modeled as a normalized Gaussian centered near
	the two-body estimate in Eq.~\eqref{eq:stripped_energy_center}.  Its width is
	chosen as
	\begin{equation}
		\Delta E_X
		=
		\max\left[
		0.15~{\rm keV},\,
		0.15\,\max\!\left(E_\mu^{(0)},1~{\rm keV}\right)
		\right].
		\label{eq:spectrum_width_model}
	\end{equation}
	This reference width represents pulse bandwidth, initial stuck-ion motion,
	target-medium broadening, and post-stripping collisions at the level of the
	present effective spectrum.  Its impact is tested in
	Sec.~\ref{subsec:robustness}.
	
	For the transport part of the calculation, the effective capture cross section
	and stopping power are parametrized as
	\begin{align}
		\sigma_{\rm cap}^{\rm eff}(E)
		&=
		\sigma_{\rm cap}^{\rm ref}
		\left(
		\frac{E_{\rm cap}^{\rm ref}+E_0}{E+E_0}
		\right)^{p_{\rm cap}},
		\label{eq:sigma_cap_model}
		\\
		S_{\rm eff}(E)
		&=
		\phi S_{\rm ref}
		\left(
		\frac{E_{\rm stop}^{\rm ref}+E_0}{E+E_0}
		\right)^{p_{\rm stop}} .
		\label{eq:stopping_model}
	\end{align}
	The first expression is a monotonic low-energy-enhancement ansatz for the
	effective capture cross section entering the kinetic-theory collision rate
	\(n\sigma v\) \cite{chapman1970mathematical}.  The second uses the standard
	stopping-power concept \(S=-dE/dx\), whose microscopic basis is the
	charged-particle energy-loss theory of Bethe \cite{bethe1930theorie}.  The
	normalizations \(\sigma_{\rm cap}^{\rm ref}\) and \(S_{\rm ref}\) are scanned
	to identify the transport window required for recycling.
	
	We use
	\begin{equation}
		\begin{aligned}
			E_{\rm cap}^{\rm ref}
			&=
			E_{\rm stop}^{\rm ref}
			=
			5~{\rm keV},
			&
			E_0
			&=
			0.10~{\rm keV},
			\\
			p_{\rm cap}
			&=
			p_{\rm stop}
			=
			0.50 .
		\end{aligned}
	\end{equation}
	The reference amplitudes \(\sigma_{\rm cap}^{\rm ref}\) and \(S_{\rm ref}\)
	are varied across the benchmark scenarios and scans.
	
	The effective molecular-formation rate used in the numerical calculation is
	\begin{equation}
		\lambda_{dt\mu}^{\rm eff}
		=
		\phi\,4c_Dc_T
		\left(
		\lambda_{dt\mu}^{\rm base}
		+
		\lambda_{dt\mu}^{\rm res}
		\right).
		\label{eq:dtmu_eff_numerical}
	\end{equation}
	The factor \(\phi\) gives the leading density scaling, and \(4c_Dc_T\) is a
	normalized D--T pair-availability factor equal to unity for an equal D--T
	mixture.  In realistic \(\mu\)CF kinetics, the molecular-formation rate
	depends on density, temperature, isotope composition, target phase, and
	hyperfine populations
	\cite{Ackerbauer:1999,ponomarev1990muon,froelich1992muon,Petitjean:1992iq,Yamashita:2022rtu}.
	
	For the ordinary molecular-formation contribution we use
	\begin{equation}
		\lambda_{dt\mu}^{\rm base}
		=
		1.1\times10^8~{\rm s}^{-1}.
	\end{equation}
	The second term, \(\lambda_{dt\mu}^{\rm res}\), represents an additional
	resonant fast channel.  It is motivated by epithermal resonant \(dt\mu\)
	formation and by direct atomic-beam measurements of resonant \(d\mu t\)
	formation \cite{Cohen:1985,Fujiwara:2000PRL}.  The largest observed resonance
	peak corresponds to a formation-rate scale of order \(10^{10}~{\rm s}^{-1}\),
	which motivates \(\lambda_{dt\mu}^{\rm res}=10^{10}~{\rm s}^{-1}\) in the
	resonant and optimistic scenarios \cite{Fujiwara:2000PRL}.  Recent kinetic
	and spectroscopic studies of resonance-state muonic molecules further support
	treating resonant pathways as an effective fast channel in \(\mu\)CF cycle
	modeling \cite{Yamashita:2022rtu,yamashita2025radiative,toyama2026direct}.
	
	The four benchmark scenarios are summarized in
	Table~\ref{tab:benchmark_scenarios}.  They are benchmark transport regimes
	rather than experimental configurations.  The set includes an
	escape-dominated conservative case, a baseline case with moderate capture and
	slowing, a baseline case supplemented by a resonant \(dt\mu\) fast channel,
	and an optimistic case with stronger capture, stronger slowing, longer
	confinement, and reduced atomic-stage losses.
	
	\begin{table*}[htbp]
		\caption{
			Benchmark transport scenarios used in the post-stripping rate-network
			calculation.  The quantities \(\sigma_{\rm cap}^{\rm ref}\) and \(S_{\rm ref}\)
			are the reference amplitudes in Eqs.~\eqref{eq:sigma_cap_model} and
			\eqref{eq:stopping_model}.  The parameter \(L_{\rm eff}\) controls the escape
			time in Eq.~\eqref{eq:lambda_escape}, and \(\kappa_{\rm esc}\) is the
			corresponding escape-suppression factor.  The rates
			\(\lambda_{\rm atom}^{\rm loss}\) and \(\lambda_{\rm atom}^{\rm esc}\)
			describe loss and escape after capture into the \(d\mu/t\mu\) atomic stage,
			while \(\lambda_{dt\mu}^{\rm res}\) parametrizes an effective resonant fast
			channel for molecular formation.
		}
		\label{tab:benchmark_scenarios}
		\begin{ruledtabular}
			\begin{tabular}{lccccccc}
				Scenario
				& \(L_{\rm eff}\) (cm)
				& \(\kappa_{\rm esc}\)
				& \(\sigma_{\rm cap}^{\rm ref}\) (cm\(^{2}\))
				& \(S_{\rm ref}\) (keV/cm)
				& \(\lambda_{\rm atom}^{\rm loss}\) (s\(^{-1}\))
				& \(\lambda_{\rm atom}^{\rm esc}\) (s\(^{-1}\))
				& \(\lambda_{dt\mu}^{\rm res}\) (s\(^{-1}\))
				\\
				\hline
				Conservative
				& \(3.0\times10^{-2}\)
				& \(1.0\)
				& \(3.0\times10^{-22}\)
				& \(20\)
				& \(1.0\times10^{8}\)
				& \(1.0\times10^{7}\)
				& \(0\)
				\\
				Baseline
				& \(1.0\times10^{-1}\)
				& \(1.0\)
				& \(1.0\times10^{-21}\)
				& \(100\)
				& \(1.0\times10^{7}\)
				& \(1.0\times10^{6}\)
				& \(0\)
				\\
				Baseline + resonant
				& \(1.0\times10^{-1}\)
				& \(1.0\)
				& \(1.0\times10^{-21}\)
				& \(100\)
				& \(1.0\times10^{7}\)
				& \(1.0\times10^{6}\)
				& \(1.0\times10^{10}\)
				\\
				Optimistic
				& \(3.0\times10^{-1}\)
				& \(0.3\)
				& \(3.0\times10^{-21}\)
				& \(300\)
				& \(0\)
				& \(0\)
				& \(1.0\times10^{10}\)
			\end{tabular}
		\end{ruledtabular}
	\end{table*}
	
	The standard calculation uses 860 energy-grid points from
	\(E_{\rm min}=0.01~{\rm keV}\) to \(E_{\rm max}=150~{\rm keV}\).  The
	spectrum averages in Eqs.~\eqref{eq:etaX_integral}--\eqref{eq:Pform_integral}
	are evaluated by trapezoidal integration.  For each scenario in
	Table~\ref{tab:benchmark_scenarios}, the calculation returns \(\eta_X\), the
	post-stripping loss budget, \(R_X=f_XP_X\eta_X\), the effective sticking
	probability \(\omega_S^{\rm eff}\), and the cycle-yield gain \(G_N\).
	
	Several consistency checks are imposed.  First, the absorbing probabilities
	must satisfy the budget normalization in Eq.~\eqref{eq:budget_sum}.  Second,
	the limits \(P_X=0\) and \(f_X=0\) must reproduce the collision-only result.
	Third, the collision-only reference ratio must satisfy
	\begin{equation}
		\frac{\omega_{S,{\rm col}}^{\rm eff}}{\omega_S^0}
		=
		1-R_{\rm col}
		=
		0.65.
	\end{equation}
	Finally, the stability of \(\eta_X\) is checked on coarser and finer energy
	grids.  These validation and robustness tests are reported together with the
	numerical results in Sec.~\ref{sec:results}.

	\section{Results and robustness}
	\label{sec:results}
	
	We now present the rate-network results.  The discussion follows the logic of
	the criterion developed above: first the benchmark recycling probabilities,
	then the associated loss channels, the transport window, the overlap
	constraint, and finally the numerical robustness tests.
	
	\subsection{Benchmark scenarios}
	\label{subsec:benchmark_results}
	
\begin{figure*}[htbp]
		\centering
		\begin{minipage}{0.48\textwidth}
			\centering
			\includegraphics[width=\linewidth]{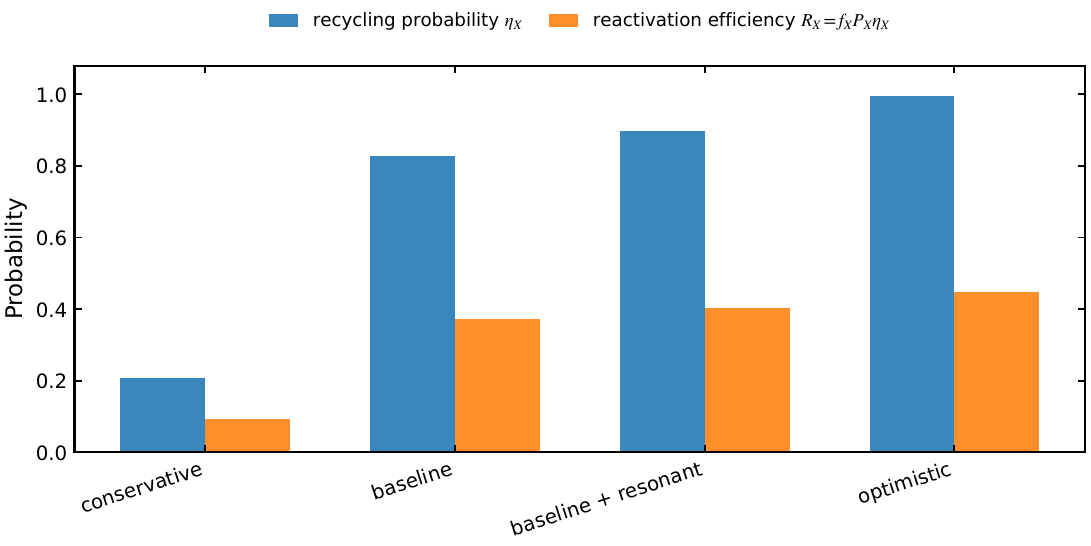}
			\vspace{-2mm}
			\centerline{(a)}
		\end{minipage}
		\hfill
		\begin{minipage}{0.48\textwidth}
			\centering
			\includegraphics[width=\linewidth]{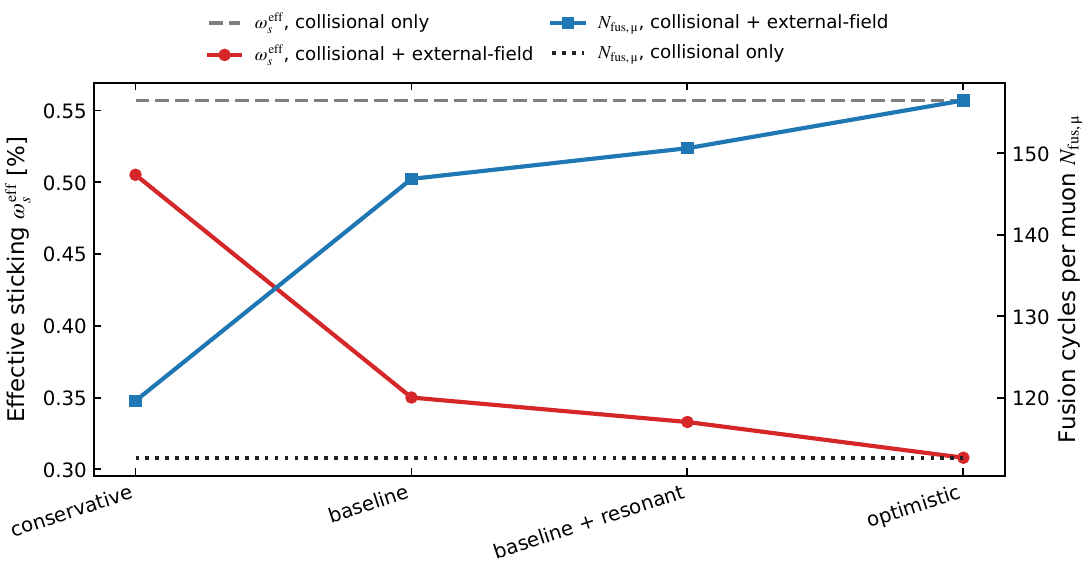}
			\vspace{-2mm}
			\centerline{(b)}
		\end{minipage}
		\caption{
			Benchmark scenario results.  Panel (a) shows the post-stripping recycling
			probability \(\eta_X\) and the effective external reactivation probability
			\(R_X=f_XP_X\eta_X\).  Panel (b) shows the corresponding effective sticking
			probability and the average number of fusion reactions per muon.  The
			collision-only reference values are
			\(\omega_{S,{\rm col}}^{\rm eff}=0.557\%\) and
			\(N_{\rm fus,\mu}=112.6\).
		}
		\label{fig:scenario_results}
	\end{figure*}

	Figure~\ref{fig:scenario_results} and Table~\ref{tab:benchmark_results}
	summarize the four benchmark scenarios defined in
	Table~\ref{tab:benchmark_scenarios}.  The collision-only reference gives
	\begin{equation}
		\omega_{S,{\rm col}}^{\rm eff}
		=
		0.557\%,
		\qquad
		N_{\rm fus,\mu}
		=
		112.6 .
	\end{equation}
	For the reference photon field, the fluence model gives \(P_X=0.449\).  In
	this subsection we set \(f_X=1\).
	
	\begin{table*}[htbp]
		\caption{
			Benchmark scenario results for the reference photon field,
			\(P_X=0.449\), and \(f_X=1\).  The gain factor is defined with respect to the
			collision-only reference case.
		}
		\label{tab:benchmark_results}
		\begin{ruledtabular}
			\begin{tabular}{lccccc}
				Scenario
				& \(\eta_X\)
				& \(R_X\)
				& \(\omega_S^{\rm eff}\) (\%)
				& \(N_{\rm fus,\mu}\)
				& \(G_N\)
				\\
				\hline
				Conservative
				& 0.207
				& 0.093
				& 0.505
				& 119.6
				& 1.062
				\\
				Baseline
				& 0.828
				& 0.372
				& 0.350
				& 146.9
				& 1.304
				\\
				Baseline + resonant
				& 0.897
				& 0.402
				& 0.333
				& 150.6
				& 1.338
				\\
				Optimistic
				& 0.996
				& 0.447
				& 0.308
				& 156.5
				& 1.390
			\end{tabular}
		\end{ruledtabular}
	\end{table*}
	
	The benchmark scenarios show a clear hierarchy.  The conservative case gives only
	\(\eta_X=0.207\) and \(G_N=1.062\), because many stripped muons are lost before
	they re-enter the catalytic cycle.  The baseline case already reaches
	\(\eta_X=0.828\) and \(G_N=1.304\).  Adding the resonant \(dt\mu\) channel
	raises the gain to \(G_N=1.338\), while the optimistic scenario gives
	\(\eta_X=0.996\) and \(G_N=1.390\).
	
	In the best-performing benchmark,
	\[
	N_{\rm fus,\mu}
	=
	156.5 ,
	\]
	compared with \(112.6\) in the collision-only reference case.  The external
	channel therefore adds about \(43.9\) fusion reactions per muon for the
	adopted reference photon field.  For the same \(P_X\) and \(f_X=1\), the
	formal upper limit \(\eta_X=1\) would give
	\(N_{\rm fus,\mu}\simeq156.7\).  Thus the optimistic benchmark nearly
	saturates the transport limit for the chosen photon-field and overlap
	parameters.
	
	Since the photon-field parameters are identical in all four scenarios, the
	variation in \(G_N\) is driven by the post-stripping recycling probability
	\(\eta_X\), not by the microscopic stripping probability \(P_X\).  This is the
	central practical result of the benchmark scan: the useful control variable is
	the product \(f_XP_X\eta_X\).
	
	\subsection{Post-stripping loss budget}
	\label{subsec:loss_budget}
	
\begin{figure}[htbp]
		\centering
		\includegraphics[width=0.49\textwidth]{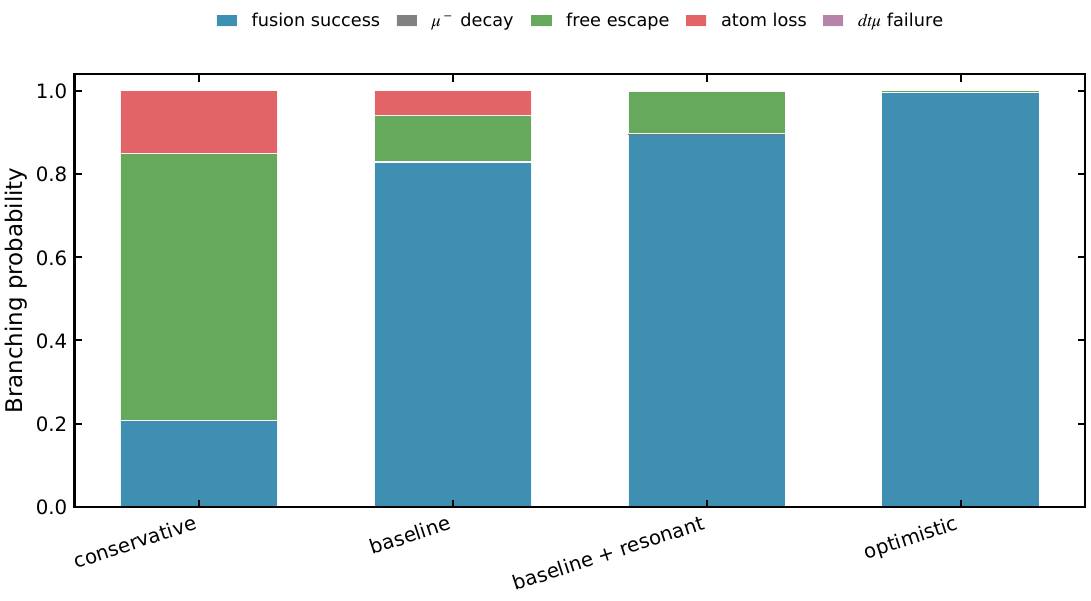}
		\caption{
			Post-stripping loss budget for the four benchmark scenarios.  The fusion
			component is the recycling probability \(\eta_X\).  The remaining
			components show the principal loss channels.  The \(dt\mu\)-failure
			component corresponds to \(P_{\rm form}\) and is below the visual
			precision of the plotted budgets.
		}
		\label{fig:loss_budget}
	\end{figure}

	The origin of the benchmark hierarchy is shown in
	Fig.~\ref{fig:loss_budget} and Table~\ref{tab:loss_budget}.  The full
	post-stripping budget contains the five absorbing components defined in
	Eqs.~\eqref{eq:absorbing_probabilities}--\eqref{eq:budget_sum}.  In the
	benchmarks, the residual molecular-channel failure \(P_{\rm form}\) is below
	the tabulated precision.  Table~\ref{tab:loss_budget} therefore lists the
	dominant components: successful recycling, free escape, atomic-stage loss, and
	muon decay.  The normalization check still includes \(P_{\rm form}\).
	
	\begin{table}[htbp]
		\caption{
			Post-stripping probability budget for the benchmark scenarios.  The listed
			entries are the dominant components of the full budget.  The residual
			molecular-channel failure \(P_{\rm form}\) is retained in the normalization
			check but is negligible at the precision shown here.
		}
		\label{tab:loss_budget}
		\begin{ruledtabular}
			\begin{tabular}{lcccc}
				Scenario
				& \(\eta_X\)
				& \(P_{\rm esc}\)
				& \(P_{\rm atom}\)
				& \(P_{\rm dec}\)
				\\
				\hline
				Conservative
				& 0.207
				& 0.641
				& 0.151
				& \(7.2\times10^{-4}\)
				\\
				Baseline
				& 0.828
				& 0.109
				& 0.060
				& \(2.8\times10^{-3}\)
				\\
				Baseline + resonant
				& 0.897
				& 0.103
				& \(7.0\times10^{-5}\)
				& \(4.5\times10^{-4}\)
				\\
				Optimistic
				& 0.996
				& \(5.1\times10^{-5}\)
				& 0
				& \(8.9\times10^{-5}\)
			\end{tabular}
		\end{ruledtabular}
	\end{table}
	
	The conservative scenario is escape dominated: about \(64\%\) of stripped
	muons leave the active region before recycling.  This explains why the cycle
	gain remains small even when stripping occurs.  In the baseline scenario,
	escape and atomic-stage losses are both reduced, and more than \(80\%\) of
	stripped muons return to fusion.  The resonant fast channel mainly suppresses
	atomic-stage loss, reducing it from about \(6\%\) to below \(10^{-3}\).  In the
	optimistic scenario, both escape and atomic-stage losses are strongly
	suppressed, so nearly every stripped muon is recycled.
	
	The rate network therefore identifies the main bottleneck: external
	reactivation is limited before the final \(dt\mu\) fusion step.  The stripped
	muon must first remain confined, slow down, and enter the fusion-active
	molecular channel.
	
	\subsection{Transport window}
	\label{subsec:transport_window}
	
\begin{figure*}[htbp]
		\centering
		\begin{minipage}{0.48\textwidth}
			\centering
			\includegraphics[width=\linewidth]{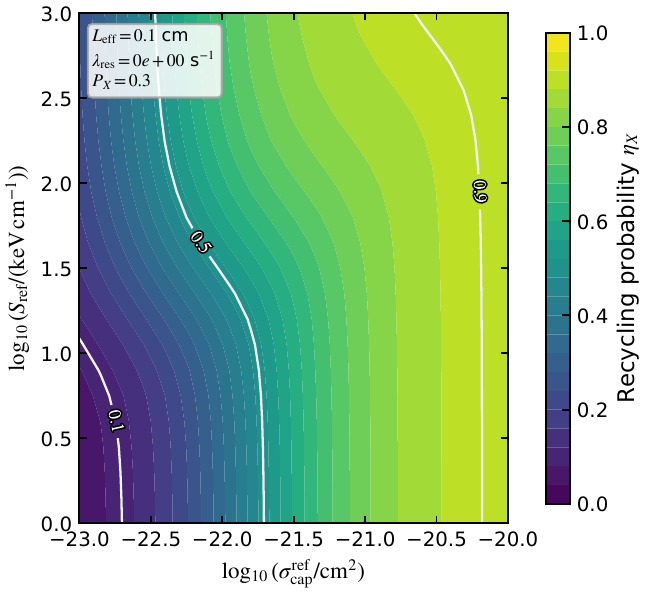}
			\vspace{-2mm}
			\centerline{(a)}
		\end{minipage}
		\hfill
		\begin{minipage}{0.48\textwidth}
			\centering
			\includegraphics[width=\linewidth]{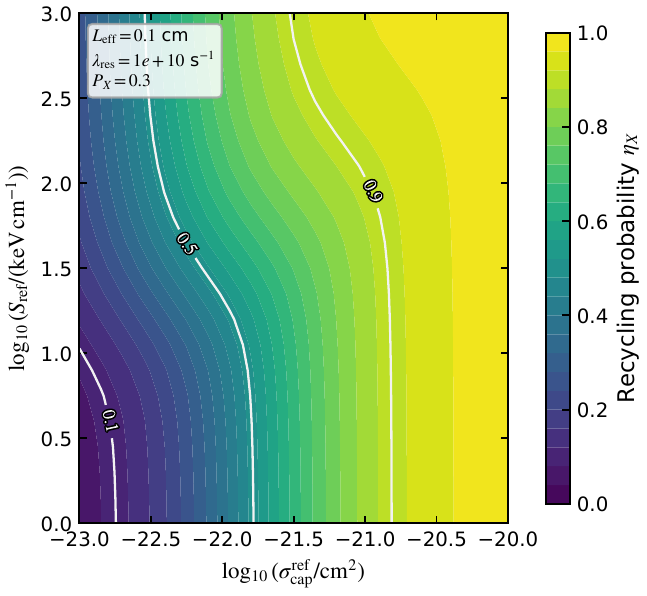}
			\vspace{-2mm}
			\centerline{(b)}
		\end{minipage}
		\caption{
			Transport-window maps for the post-stripping recycling probability
			\(\eta_X\) at \(L_{\rm eff}=0.1~{\rm cm}\) and \(P_X=0.3\).  The horizontal
			axis is the effective capture amplitude \(\sigma_{\rm cap}^{\rm ref}\), and
			the vertical axis is the effective stopping amplitude \(S_{\rm ref}\).
			Panel (a) shows the case without a resonant \(dt\mu\) fast channel.
			Panel (b) includes \(\lambda_{dt\mu}^{\rm res}=10^{10}~{\rm s}^{-1}\).
		}
		\label{fig:transport_window}
	\end{figure*}

	We next scan the effective capture amplitude
	\(\sigma_{\rm cap}^{\rm ref}\) and the effective stopping amplitude
	\(S_{\rm ref}\).  Figure~\ref{fig:transport_window} shows the resulting
	transport-window maps for \(L_{\rm eff}=0.1~{\rm cm}\) and \(P_X=0.3\), with
	and without an effective resonant fast channel.  Representative values are
	listed in Table~\ref{tab:transport_window_summary}.
	
	\begin{table*}[htbp]
		\caption{
			Representative values from the transport-window scan at
			\(L_{\rm eff}=0.1~{\rm cm}\) and \(P_X=0.3\).
		}
		\label{tab:transport_window_summary}
		\begin{ruledtabular}
			\begin{tabular}{lcc}
				Case
				& \(\eta_X\)
				& Physical interpretation
				\\
				\hline
				Weak capture/slowing, no resonance
				& 0.053
				& escape dominated
				\\
				Maximum in scan, no resonance
				& 0.913
				& high recycling
				\\
				Maximum in scan, resonant channel
				& 0.988
				& near-complete recycling
			\end{tabular}
		\end{ruledtabular}
	\end{table*}
	
	At small \(\sigma_{\rm cap}^{\rm ref}\) and small \(S_{\rm ref}\), the muon
	remains energetic for too long and often escapes before entering the
	\(d\mu/t\mu\) atomic stage.  A representative point in this region gives
	\(\eta_X\simeq0.053\), with free escape accounting for more than \(94\%\) of
	the post-stripping budget.  Increasing capture or slowing moves the system
	into a high-recycling window.  In the displayed scan, the maximum recycling
	probability is \(\eta_X\simeq0.913\) without the resonant channel and
	\(\eta_X\simeq0.988\) with
	\(\lambda_{dt\mu}^{\rm res}=10^{10}~{\rm s}^{-1}\).
	
	The resonant channel enlarges the high-\(\eta_X\) region by reducing the time
	spent in the atomic stage.  It does not remove the escape bottleneck when
	capture and stopping are both weak.  Resonant \(dt\mu\) formation is therefore
	most effective after free-muon transport has already been brought into the
	recycling window.
	
	\subsection{No-go criterion and overlap sensitivity}
	\label{subsec:overlap_nogo}
	\label{subsec:nogo_overlap}
	
\begin{figure}[htbp]
		\centering
		\includegraphics[width=0.49\textwidth]{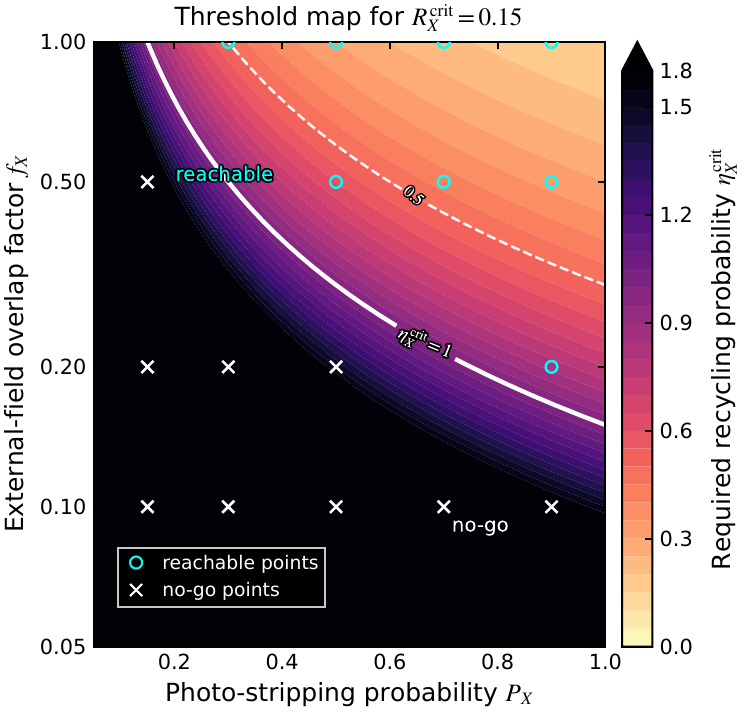}
		\caption{
			Probability-level no-go criterion for a target external reactivation
			probability \(R_X^{\rm crit}=0.15\).  The plotted quantity is the required
			post-stripping recycling probability
			\(\eta_X^{\rm crit}=R_X^{\rm crit}/(f_XP_X)\).  Regions with
			\(\eta_X^{\rm crit}>1\) are excluded independently of the detailed
			transport model.
		}
		\label{fig:nogo_threshold}
	\end{figure}

\begin{figure}[htbp]
		\centering
		\includegraphics[width=0.49\textwidth]{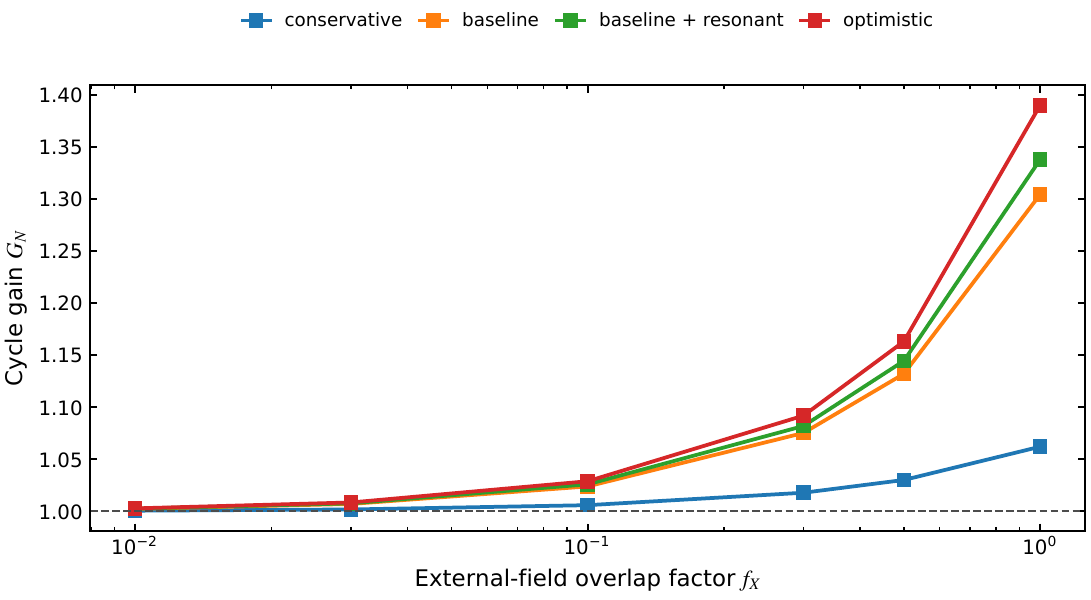}
		\caption{
			Sensitivity of the catalytic-yield gain \(G_N\) to the overlap factor
			\(f_X\) for the benchmark scenarios.  Even favorable post-stripping
			transport gives little improvement if the external field overlaps with
			only a small fraction of the residual stuck \((\alpha\mu)\) population.
		}
		\label{fig:fx_sensitivity}
	\end{figure}

	The transport-window maps give the attainable \(\eta_X\).  The no-go
	criterion gives the required \(\eta_X\).  Figure~\ref{fig:nogo_threshold}
	shows the required recycling probability for a target external reactivation
	probability, and Fig.~\ref{fig:fx_sensitivity} shows the dependence of the
	cycle gain on the overlap factor \(f_X\). 
	
	For the illustrative target value \(R_X^{\rm crit}=0.15\),
	\begin{equation}
		\eta_X^{\rm crit}
		=
		\frac{0.15}{f_XP_X}.
	\end{equation}
	Representative values are listed in Table~\ref{tab:nogo_examples}.  The
	constraint is controlled by the product \(f_XP_X\): even excellent
	post-stripping recycling cannot help if the external field samples too small
	a fraction of the residual stuck population.
	
	\begin{table}[htbp]
		\caption{
			Representative no-go thresholds for \(R_X^{\rm crit}=0.15\).  Values with
			\(\eta_X^{\rm crit}>1\) are excluded at the probability level.
		}
		\label{tab:nogo_examples}
		\begin{ruledtabular}
			\begin{tabular}{ccc}
				\(P_X\) & \(f_X\) & \(\eta_X^{\rm crit}\) \\
				\hline
				0.1 & 1.0 & 1.50 \\
				0.3 & 1.0 & 0.50 \\
				0.5 & 1.0 & 0.30 \\
				0.5 & 0.1 & 3.00
			\end{tabular}
		\end{ruledtabular}
	\end{table}
	
	The same limitation appears in Fig.~\ref{fig:fx_sensitivity}.  In the
	baseline scenario, reducing \(f_X\) from 1 to 0.1 lowers \(R_X\) from 0.372 to
	0.037 and reduces the gain from \(G_N=1.304\) to \(G_N=1.024\).  In the
	optimistic scenario, the same reduction lowers the gain from \(G_N=1.390\) to
	\(G_N=1.029\).  Thus, nearly ideal post-stripping transport is not sufficient
	unless the residual stuck population overlaps efficiently with the external
	field.
	
	\subsection{Numerical and model robustness}
	\label{subsec:robustness}
	
\begin{figure*}[t]
		\centering
		\begin{minipage}{0.48\textwidth}
			\centering
			\includegraphics[width=\linewidth]{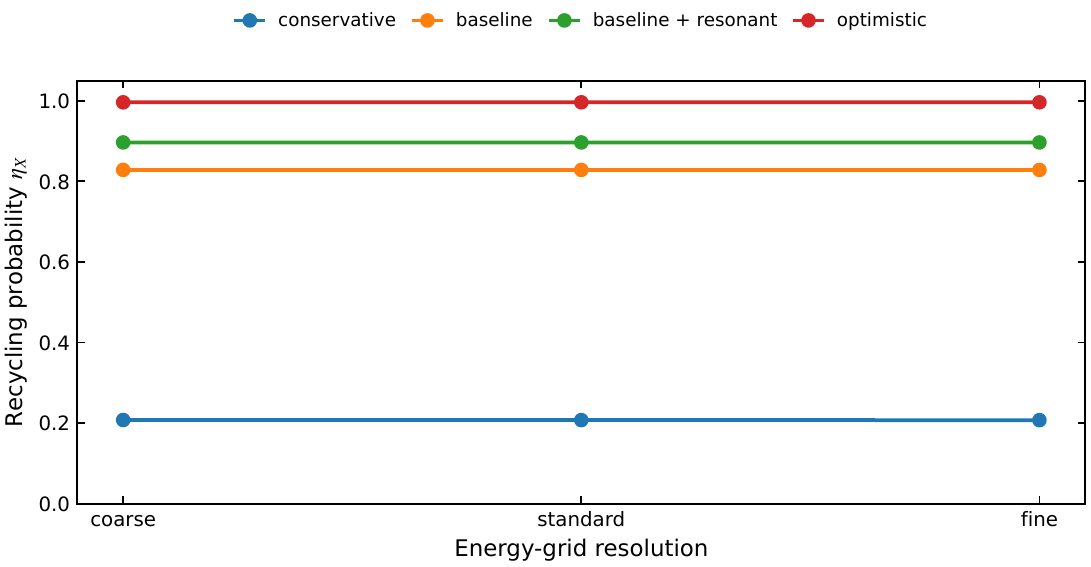}
			\vspace{-2mm}
			\centerline{(a)}
		\end{minipage}
		\hfill
		\begin{minipage}{0.48\textwidth}
			\centering
			\includegraphics[width=\linewidth]{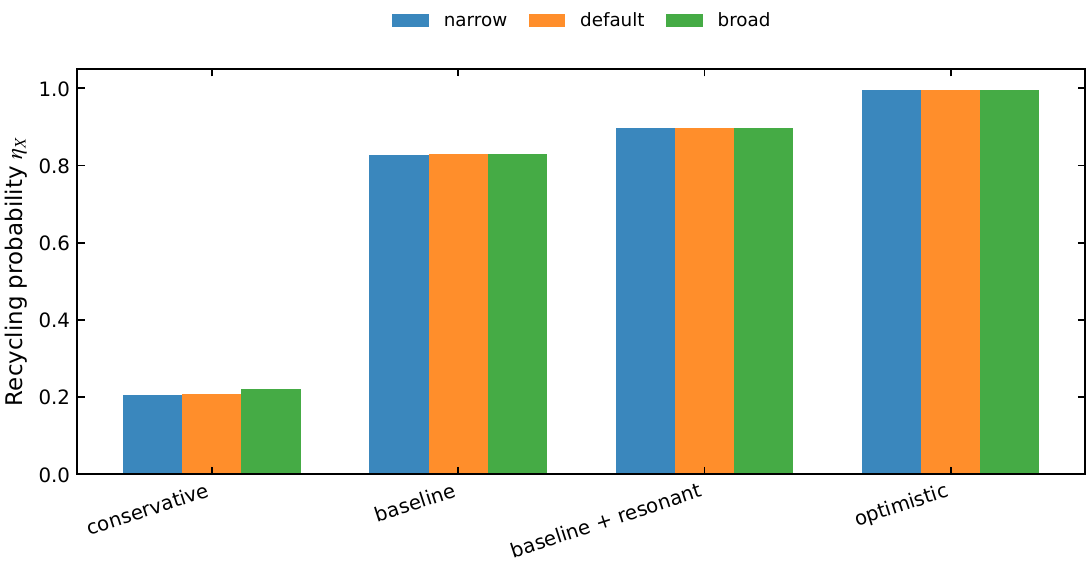}
			\vspace{-2mm}
			\centerline{(b)}
		\end{minipage}
		\caption{
			Robustness checks.  Panel (a) shows the convergence of the post-stripping
			recycling probability \(\eta_X\) with respect to the stripped-muon energy
			grid.  Panel (b) compares \(\eta_X\) obtained with narrow, default, and broad
			stripped-muon spectra.  The benchmark hierarchy is stable under both tests.
		}
		\label{fig:robustness_checks}
	\end{figure*}

	We test the numerical stability of the energy integration and the dependence
	on the assumed stripped-muon spectrum.  The results are shown in
	Fig.~\ref{fig:robustness_checks}, and the numerical ranges are summarized in
	Table~\ref{tab:robustness_summary}.
	
	\begin{table*}[htbp]
		\caption{
			Robustness summary.  The grid test compares coarse, standard, and fine energy
			grids.  The spectrum test compares narrow, default, and broad effective
			stripped-muon spectra.
		}
		\label{tab:robustness_summary}
		\begin{ruledtabular}
			\begin{tabular}{lcc}
				Scenario
				& Grid range of \(\eta_X\)
				& Spectrum-model range of \(\eta_X\)
				\\
				\hline
				Conservative
				& 0.207--0.208
				& 0.206--0.221
				\\
				Baseline
				& 0.8284--0.8285
				& 0.828--0.829
				\\
				Baseline + resonant
				& 0.8966--0.8967
				& 0.896--0.897
				\\
				Optimistic
				& 0.996258--0.996259
				& 0.9962--0.9963
			\end{tabular}
		\end{ruledtabular}
	\end{table*}
	
	For the baseline scenario, the coarse, standard, and fine grids give
	\(\eta_X=0.82847\), \(0.82845\), and \(0.82844\), respectively.  These values
	all round to \(\eta_X=0.828\) at the precision used for the main results.  The
	other scenarios show the same grid stability.  The spectrum test is also
	stable: in the baseline scenario, \(\eta_X\) varies only from 0.828 to 0.829
	when the spectrum is changed from narrow to broad.  The conservative scenario
	is more sensitive because slowing and escape compete more sharply, but the
	scenario hierarchy remains unchanged.
	
	The validation limits are satisfied as well.  Setting \(P_X=0\) or \(f_X=0\)
	reproduces the collision-only result.  Removing all reactivation gives
	\(\omega_S^{\rm eff}/\omega_S^0=1\), while the formal \(R_X=1\) limit gives
	\(\omega_S^{\rm eff}=0\).  The post-stripping budgets satisfy
	Eq.~\eqref{eq:budget_sum} in all benchmark scenarios.  These checks confirm
	that the numerical implementation preserves the probability structure of the
	rate-network criterion.
	
	\subsection{Broader uncertainty scan}
	\label{subsec:uncertainty_scan}
	
	As a final diagnostic, we vary the external-field and transport parameters
	simultaneously over representative ranges.  The scan is not used to assign
	statistical uncertainties; it tests whether the relation between the net
	external reactivation probability \(R_X\) and the cycle-yield gain \(G_N\)
	remains stable under combined parameter variations.
	
	\begin{figure}[htbp]
		\centering
		\includegraphics[width=0.49\textwidth]{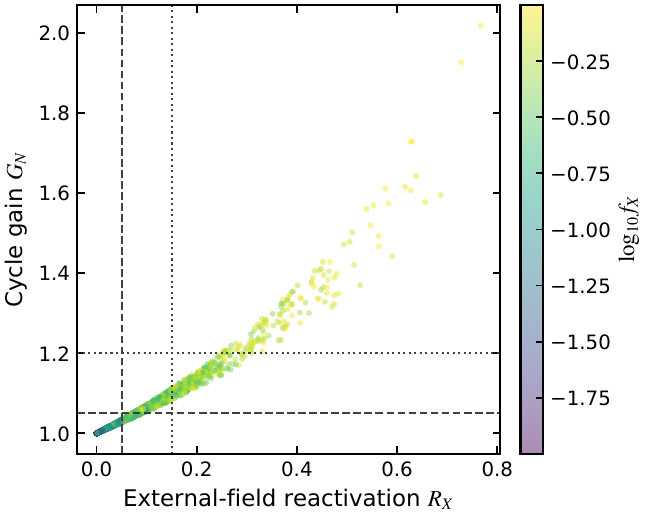}
		\caption{
			Broader uncertainty scan showing the correlation between the external
			reactivation probability \(R_X\) and the catalytic-yield gain \(G_N\).
			The scan is used as a diagnostic test of the rate-network criterion under
			simultaneous variations of the effective input parameters.
		}
		\label{fig:uncertainty_rx_gn}
	\end{figure}
	
	Figure~\ref{fig:uncertainty_rx_gn} shows the expected monotonic trend: a
	larger net external reactivation probability generally gives a larger number
	of fusions per muon.  The spread of points reflects the fact that different
	combinations of overlap, stripping, and recycling can lead to the same
	\(R_X\).  The global correlation confirms the benchmark conclusion.  The
	relevant control variable for catalytic improvement is not the stripping
	probability alone, but the net probability that a stripped muon is returned to
	the catalytic cycle.

	\section{Conclusion and outlook}
	\label{sec:conclusion}
	
	We have formulated a rate-network criterion for external-field-assisted muon
	reactivation in deuterium--tritium muon-catalyzed fusion.  The main point is
	that an external stripping event is useful only when it is followed by
	successful recycling of the liberated muon.  The external branch is therefore
	controlled jointly by the field--population overlap, the microscopic stripping
	probability, and the post-stripping recycling probability.  This separation
	also gives a simple probability-level no-go test: if the recycling probability
	required for a target gain exceeds unity, that parameter region is excluded
	before any detailed transport calculation is needed.
	
	The numerical scans show that post-stripping transport is the decisive
	bottleneck.  In the escape-dominated benchmark, external stripping gives only
	a modest increase in the cycle yield because most stripped muons leave the
	active region before entering the \(d\mu/t\mu\) atomic stage.  In the baseline
	and resonant benchmarks, improved capture, slowing, and molecular formation
	increase the recycling probability and reduce the effective sticking loss.
	For the reference inputs used here, the best-performing benchmark increases
	the cycle yield from \(N_{\rm fus,\mu}=112.6\) in the collision-only case to
	\(N_{\rm fus,\mu}=156.5\).  This value is close to the transport upper limit
	for the adopted photon-field and overlap parameters.  The resonant \(dt\mu\)
	channel suppresses atomic-stage loss and broadens the high-recycling window,
	but it cannot compensate for prompt free escape.  The overlap factor is also
	essential: even efficient recycling produces little gain if the external field
	samples only a small fraction of the residual stuck population.
	
	The analysis identifies the simultaneous requirements for a useful
	external-reactivation scheme.  The external field must provide a sizable
	stripping probability, overlap with the residual \((\alpha\mu)^+\) population
	at the right time and position, and produce stripped \(\mu^-\) that remain
	confined long enough to slow down, be captured, and return to the \(dt\mu\)
	fusion channel.  The key microscopic inputs still needed are the
	\(\alpha\mu\) photostripping cross section, the stripped-muon spectrum,
	capture and stopping of stripped \(\mu^-\) in D--T matter, and state-resolved
	resonant molecular-formation rates.  These quantities can be constrained by
	few-body bound--continuum calculations and targeted component tests.
	
	Existing and developing muon facilities, including J-PARC MUSE and
	prospective HIAF-based muon-source concepts, can constrain the muon-side
	transport and recycling processes
	\cite{Shimomura:2024MUSE,Xu:2025HIAFMuon,Cai:2024HighIntensityMuon}.  Modern
	high-repetition-rate x-ray and extreme-light facilities, such as European
	XFEL, LCLS-II, SHINE, and ELI, provide relevant photon-side capabilities for
	future synchronized stripping studies
	\cite{Decking:2020EuropeanXFEL,Rini:2023LCLSII,Liu:2023SHINEFELII,ELI:2026ERIC}.
	The rate-network framework developed here provides a quantitative basis for
	testing whether such external-field concepts can reduce residual alpha
	sticking in practical \(\mu\)CF conditions.

	\begin{acknowledgments}
		We thank Professors Masayasu Kamimura and Yasushi Kino for valuable discussions
		and insightful suggestions on muon-catalyzed fusion dynamics. This work has
		been supported by the National Natural Science Foundation of China (Grant No.
		12547118), the Research Program of State Key Laboratory of Heavy Ion Science
		and Technology, Institute of Modern Physics, Chinese Academy of Sciences
		(Grant No. HIST2025CS08), and the National Key R\&D Program of China (Grant
		Nos. 2024YFE0109800 and 2024YFE0109802).
	\end{acknowledgments}
	

	 \bibliographystyle{apsrev4-2}
	\bibliography{refs}
	
\end{document}